# MAPPING THE SPATIAL DISTRIBUTION OF CHARGE CARRIERS IN LaAlO$_3$/SrTiO$_3$ HETEROSTRUCTURES

M. Basletic, J.-L. Maurice, C. Carrétéro, G. Herranz, O. Copie, M. Bibes, É. Jacquet, K. Bouzehouane, S. Fusil and A. Barthélémy

Unité Mixte de Physique CNRS/Thales, associée à l'Université Paris-Sud, Route départementale 128, 91767 Palaiseau Cedex, France

Abstract

At the interface between complex insulating oxides, novel phases with interesting properties may occur, such as the metallic state reported in the LaAlO$_3$/SrTiO$_3$ system [1]. While this state has been predicted [2] and reported [3,4] to be confined at the interface, some works indicate a much broader spatial extension [5], thereby questioning its origin. Here we provide for the first time a direct determination of the carrier density profile of this system through resistance profile mappings collected in cross-section LaAlO$_3$/SrTiO$_3$ samples with a conducting-tip atomic force microscope (CT-AFM). We find that, depending upon specific growth protocols, the spatial extension of the high-mobility electron gas can be varied from hundreds of microns into SrTiO$_3$ to a few nanometers next to the LaAlO$_3$/SrTiO$_3$ interface. Our results emphasize the potential of CT-AFM as a novel tool to characterize complex oxide interfaces and provide us with a definitive and conclusive way to reconcile the body of experimental data in this system

Due to the richness of their intrinsic properties oxide materials can bring novel functionalities to electronics. Among all oxides, SrTiO$_3$ (STO) is particularly appealing because of its multifunctional character. It is a wide band gap semiconductor (3.2 eV) that becomes a high mobility metal [6] or even superconducting [7] upon oxygen vacancy doping. Stoichiometric STO is also at the fringe of ferroelectricity and its large dielectric permittivity, tunability and low microwave loss, makes it a good candidate for tunable microwave devices [8].

The potential of oxides is further enlarged by the fascinating properties of their interfaces, epitomized by the observation of ferromagnetism at the interface between two antiferromagnets [9] or of the quantum Hall effect in two-dimensional (2D) ZnO-based structures [10]. Recently, an electron gas with high electron mobility at low temperature (in the $10^4$ cm$^2$/Vs range) has been reported in heterostructures combining STO with LaAlO$_3$ (LAO), another band insulator [1]. Since its discovery by Ohtomo and Hwang [1], this unusual metallic state has been the object of many studies [2,3,4,5,11,12,13,14,15,16,17,18]. Recent results include a two-dimensional superconducting behaviour at the LAO/STO interface below 200 mK [19], and indications of a ferromagnetic state below ~1K [20].

The arguments invoked to explain the metallic state observed in such LAO/STO samples have been dual. A first interpretation relies on the transfer of a ½ $e$ charge per unit cell at the interface, because of the polar discontinuity between positively charged LaO and neutral TiO$_2$ sub-planes [11]. In this picture a sheet carrier density $n_{sheet} \approx 3.3 \ 10^{14}$ cm$^{-2}$ is expected. However, in many cases the reported values of $n_{sheet}$ exceed this density by orders of magnitude [1,5,17]. This leads to another interpretation relying on the creation of oxygen vacancies in the STO substrate during the growth of the LAO film, which dopes the STO with electrons and makes it a high-mobility metal, as is well-known to occur in oxygen-deficient bulk STO [6]. In this case, the thickness of the metallic region may be very large as oxygen vacancies can diffuse very rapidly over large distances in STO [21].

Efforts have been directed at the determination of the spatial distribution of the delocalized carriers in these conducting LAO/STO structures. Atomic-scale electron energy loss spectroscopy (EELS) studies have shown that there is a nanometric region confined next to the interface where a fraction of Ti with mixed 3+/4+ valence exists, indicating that some charge is transferred across the interface [11,22,23]. Nevertheless such measurements do not provide information on the localized or mobile nature of the generated charge and cannot detect the very low doping levels required for high mobility in STO [5,11,22]. Alternatively, Reyren et al. have recently demonstrated two-dimensional superconductivity at the LAO/STO interface, and estimated the conducting thickness to be at most around 10 nm at low temperature for samples deposited at $6 \cdot 10^{-5}$ mbar and annealed in 400 mbar of oxygen [19]. In contrast, Herranz et al. have deduced a thickness of ~500 μm from the analysis of Shubnikov-de Haas oscillations in LAO/STO samples grown at low pressure ($10^{-6}$ mbar, no annealing) [5]. However, in spite of the critical importance of determining the distribution of carriers in such structures [24], there have been no *direct* measurements of the thickness and local carrier density of the metallic region in LAO/STO samples.

In the present work we have addressed this issue by probing the local resistance of LAO/STO samples using a conducting-tip atomic force microscope (CT-AFM) in cross-section geometry. Through a specific calibration procedure using doped $SrTiO_3$ single crystals, we are able not only to map out the local resistance but also to measure the spatial distribution of the charge carrier density, with a resolution of a few nm. We apply the technique to samples grown and annealed in different conditions and provide for the first time a direct evaluation of the thickness and carrier density of the interfacial metallic gas at the LAO/STO interface.

Two different LAO films were grown at an oxygen pressure $P_{O2} \leq 10^{-5}$ mbar and a temperature T = 750 ºC by pulsed laser deposition (PLD) on $TiO_2$-terminated STO substrates. The growth was monitored by RHEED and intensity oscillations were consistently observed (see Figure 1c). A first LAO film of thickness 200 nm was grown at $P_{O2} = 10^{-5}$ mbar and cooled down to room temperature at the same low pressure (from now on we refer to this sample as 'non-annealed'). A second epitaxial LAO film with thickness ≈ 2 nm (5 u.c.) was grown at $P_{O2} = 10^{-6}$ mbar, and cooled down from 750°C to room temperature at $P_{O2} = 300$ mbar (we refer to this sample as 'in-situ annealed'). For this latter sample, a 200 nm-thick amorphous LAO layer was finally deposited on top of the epitaxial LAO layer at room temperature in order to track down easily the LAO/STO interface in the CT-AFM experiments.

The samples were characterized by high-resolution transmission electron microscopy (HRTEM), X-ray diffraction (XRD) and atomic force microscopy (AFM). Details on the microstructure of 'non-annealed' films can be found in Ref. [14,22,23]. As visible from the HRTEM image shown in Figure 1a, in the 'in-situ annealed' the LAO film is epitaxial and fully strained. The LAO/STO interface appears sharp from HRTEM (Fig.1a) and the surface forms flat terraces separated by one-unit-cell-high steps (Fig. 1b).

The macroscopic transport properties of the samples were first measured in standard four-probe geometry as described in Ref. [5]. Subsequently, LAO/STO cross-sectional samples were prepared as described in the Methods (see Fig. 2 for a sketch of the experiment). Resistance ($R_{CT-AFM}$) mapping profiles of cross-section samples were directly collected with CT-AFM at room temperature. In these experiments, the morphology and the resistance between the bottom electrode and the tip were mapped simultaneously by scanning the cross-sections with the AFM tip approximately parallel to the interface at a scanning frequency of 3 Hz.

We first discuss the properties of the 'non-annealed' sample. As visible from Figure 3a, four-probe resistance measurements indicate that this sample is metallic over the whole range of temperatures with low-temperature sheet resistance $R_{sheet,4K} \approx 10^{-3}$ Ω and mobility $\mu_{4K} \approx 2260$ cm$^2$V$^{-1}$s$^{-1}$. The sheet carrier density is $n_{sheet,4K} \approx 5\ 10^{17}$ cm$^{-2}$ at low temperature and $n_{sheet,300K} \approx 3\ 10^{17}$ cm$^{-2}$ at room temperature. These values are comparable with those reported elsewhere for samples grown in similar conditions [1,5,12,17]. CT-AFM measurements for this sample are shown in Fig. 3b. We observe that the resistance close to the LAO/STO interface is in the $10^6$ Ω range and increases over a region of width ~1-2 μm up to a saturation value of about $5\ 10^{10}$ Ω, and then stays at this value away from the interface. Indeed, 3×3 μm² scans $R_{CT-AFM}$ mappings measured at different distances from the interface (see inset of Fig. 5b) demonstrate that the STO substrate is conducting ($R_{CT-AFM} \approx 5\ 10^{10}$ Ω) up to the backside (~500 μm away from the interface) for the 'non-annealed' sample. This is in agreement with the conclusions drawn from our previous magnetotransport experiments in high magnetic fields for similar 'non-annealed' samples [5]. Thus, our CT-AFM measurements indicate that one can picture the 'non-annealed' LAO/STO structures as consisting of two conductive regions in parallel. One region is the bulk STO substrate, that becomes rapidly conducting when growth proceeds at low pressure, consistent with the high diffusion coefficients for oxygen vacancies in STO [21]. The other region is the interface whose resistance is lower than that of the bulk by about four orders of magnitude.

In order to get rid of the oxygen vacancies and isolate this interface conduction, we have followed the protocol indicated by Thiel et al. [4], and grown an 'in-situ annealed' sample. As visible from Fig. 4a, four-probe transport measurements indicate that this sample is also metallic, but with a much higher sheet resistance $R_{sheet,4K} \approx 10^2$ Ω. The low-temperature mobility $\mu_{4K} \approx 650$ cm$^2$V$^{-1}$s$^{-1}$, sheet carrier density $n_{sheet,4K} \approx 3\ 10^{13}$cm$^{-2}$ and room-temperature sheet carrier density $n_{sheet,300K} = 5\ 10^{14}$ cm$^{-2}$ are in the range of those reported for samples grown in similar conditions [3,4,19]. While the *macroscopic* transport properties of the 'in-situ annealed' and 'non-annealed' samples are somehow quite similar (metallic behaviour, high-mobility at low-temperature), CT-AFM experiments draw the veil onto the fundamental differences in their *local* transport properties. As shown in Fig. 4b, the CT-AFM mappings on the 'in-situ annealed' sample indicate that the bulk of the substrate is now insulating ($R_{CT-AFM} > 10^{12}$ Ω), in strong contrast with the behaviour found for the 'non-annealed' sample (Fig. 3b). Even more striking is the presence of an extremely narrow highly conducting region ($R_{CT-AFM} \approx 6\ 10^5$ Ω) just at the LAO/STO interface. To disclose the transport properties of the interface at the nanometre scale, we show a high-resolution image in Fig. 4c and plot the resistance profile across the interface in Fig. 4d. The thickness of the metallic gas is found to be only 7 nm, i.e. conduction is really confined at the interface. Taking into account that the signal is convoluted by the AFM tip we can consider this value as an upper limit of the metallic gas extension.

In order to get insight not only on the local resistance but also on the local carrier density from the CT-AFM data, we have setup a calibration procedure to obtain a direct correspondence between the carrier concentration and the CT-AFM resistance. We have first measured the carrier density and the mobility of Nb-doped STO single crystals at room temperature and prepared cross-section samples using the same preparation process as for LAO/STO samples. We note that the room temperature mobility does not vary significantly in the doping range considered ($\mu_{300K} \approx 1-5$ cm$^2$/Vs), which allows a direct conversion of $R_{CT-AFM}$ into n. As visible from Figure 5a, a systematic variation of n with $R_{CT-AFM}$ is obtained and all the data fall on the same line. The addition in this Figure of the data point corresponding to the 'non-annealed' sample (open circle) indicates that the sample preparation and measurement procedures do not depend on the dopant type (Nb ions or oxygen vacancies).

Equipped with this calibration, we can now extract the carrier density profile across the interface from the $R_{CT-AFM}$ mappings shown in Figs. 3b and 4b (see Fig. 5b). In the 'non-annealed' sample, the carrier density varies from $2\ 10^{21}$ cm$^{-3}$ close to the interface to about $5\ 10^{18}$ cm$^{-3}$ far from the interface. From the value of $n_{sheet,300K}$, we then deduce that the thickness of the metallic gas in this sample is about 600 µm, consistent with the mappings collected hundreds of microns from the interface (see inset). This calibration procedure also allows the determination of the carrier density just at the interface in the 'in-situ annealed' sample, ~7 $10^{21}$ cm$^{-3}$. Similarly, using the value of $n_{shee,t300K}$ determined previously we can give an estimate of the metallic gas thickness. This yield the remarkable value of 1 nm.

What is the origin of this interfacial metallic gas ? While for the 'in-situ annealed' sample, oxygen vacancies can in principle be ruled out, one may invoke slight La-doping at the interface driven by inter-diffusion [25] or charge transfer due to electronic reconstruction at the heteropolar LAO/STO interface as proposed by Hwang et al [1,11]. In this latter hypothesis, as previously mentioned, a charge density $n_{sheet}=3\ 10^{14}$ cm$^{-2}$ should be present, which is close to the value we obtain ($5\ 10^{14}$ cm$^{-2}$ at room temperature). Also, the spatial extension of the metallic gas within this picture, as calculated by Siemons et al [12], is compatible with our results.

In summary, we have developed a method based on conductive-tip atomic force microscopy in cross-sectional samples to map the spatial distribution of charge carriers in complex heterostructures. We have then used this technique to measure the extension of the high-mobility electron gas reported at the interface between LaAlO$_3$ and SrTiO$_3$ and the local carrier density. Our results reveal that both interface conduction and bulk conduction due to oxygen vacancies can occur depending on the growth and annealing process. More importantly, the experiments provide direct evidence for the presence of a metallic electron gas with a carrier density around $10^{21}$ cm$^{-3}$ confined at room temperature within a few nm next to the LaAlO$_3$/SrTiO$_3$ interface in annealed samples. Both the gas extension and the carrier density are compatible with a mechanism based on charge transfer as the origin of this metallic state.

METHODS

GROWTH DEPOSITION OF LAO/STO STRUCTURE

LAO thin films were grown by pulsed laser deposition (PLD) using a frequency-tripled ($\lambda$ =355 nm) Nd: yttrium aluminum garnet (YAG) laser on $TiO_2$-terminated STO substrates [Surfacenet Gmbh] with growth oxygen pressures $P_{O2} \leq 10^{-5}$ mbar at a deposition temperature of T = 750 ºC. The laser pulse rate was fixed at 2.5 Hz, and the target–substrate distance was 55 mm, resulting in a growth rate of about 0.5 Å/s.

SAMPLE PREPARATION FOR CT-AFM CHARATERIZATION OF LAO/STO STRUCTURES

Prior to CT-AFM measurements, the interface was contacted with Al/Au through the LAO layer by locally etching the LAO with accelerated Ar ions down to the interface, in a chamber equipped with a secondary ion mass spectroscopy detection system [5].

The CT-AFM is based on a Digital Instruments Nanoscope III multimode AFM, which was modified by Houzé et al. to perform local resistance measurements in the range of 100 to $10^{12}$ ohms with 5% accuracy under a bias voltage ranging from 0.1 to 10 V [26]. The LAO/STO cross-sectional samples for measurements of the local resistance with CT-AFM were cut through the contacts described above and glued together front-to-front (as for a TEM cross section) so that the contacts remained usable. The lateral surface to be studied was then subjected to a mechanical polishing with diamond coated disks, of grain size decreasing progressively to 0.5 µm, followed by a short finish using colloidal silica of 20-nm grain size. An Al/Au contact was then connected using silver paste to act as the electrode complementary to the CrPt-coated tip of the CT-AFM (see Fig. 2). The CT-AFM images were acquired with a bias voltage $V_{bias} = -1$ V between tip (grounded) and sample. In the resistance mappings obtained from CT-AFM, dark red corresponds to a local resistance of $10^5$ Ω, whereas pink is $10^{12}$ Ω, which is the highest resistance value limited by instrumental current sensitivity.

ACKNOWLEDGMENTS

G. H. acknowledges financial support from the DURSI (Generalitat de Catalunya, Spain). Financial support from PAI- France-Croatia COGITO Program No. 82/240083, Croatian MZOS Project No. 119-1191458-1023 and the French Agence Nationale de la Recherche (Project Pnano ALICANTE) is acknowledged. We thank Y. Gourdel for his help in the polishing process.

FIGURE CAPTIONS

Figure 1. **Structural and morphological characterization of LAO/STO structures.**
HRTEM (a) and AFM (b) image of the 5 unit-cell LAO film grown on STO ('in-situ annealed' sample). (c) Typical RHEED oscillations for LAO films grown on STO at $10^{-6}$ mbar at 750°C.

Figure 2. **Sketch of the CT-AFM experiment.**
As the tip is swept across the interface, the surface topography and the local resistance are mapped simultaneously. The AFM image is that of the 'non-annealed' sample.

Figure 3. **'Non-annealed' LAO/STO interface**
(a) Temperature dependence of the sheet resistance $R_{sheet}$ and (b) CT-AFM resistance mapping.

Figure 4. **'In-situ annealed' LAO/STO interface**
(a) Temperature dependence of the sheet resistance $R_{sheet}$. (b) CT-AFM resistance mapping. (c) Close-up of the $R_{CT-AFM}$ mapping around the interface. (d) Resistance profile across the LAO/STO interface extracted from (c).

Figure 5. **Nano-scale CT-AFM characterization of carrier distribution across LAO/STO interfaces**
(a) Plot of the carrier density *n* versus $R_{CT-AFM}$ measured on STO substrates doped with different concentrations of Nb ions (closed squares). The open circle denotes the *n*-$R_{CT-AFM}$ data from the 'non-annealed' STO substrate. (b) Resistance and equivalent carrier density profiles for the 'non-annealed' (blue) and 'in-situ annealed' (red) interfaces. The *n*-profiles are extracted using the calibration data displayed in (a). The inset is a CT-AFM image of the 'non-annealed' sample collected ~500 μm away from the interface.

Fig. 1

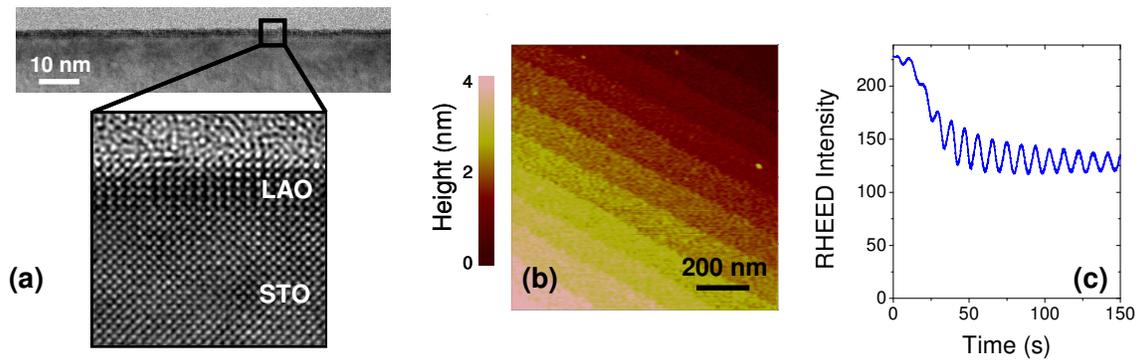

Fig. 2

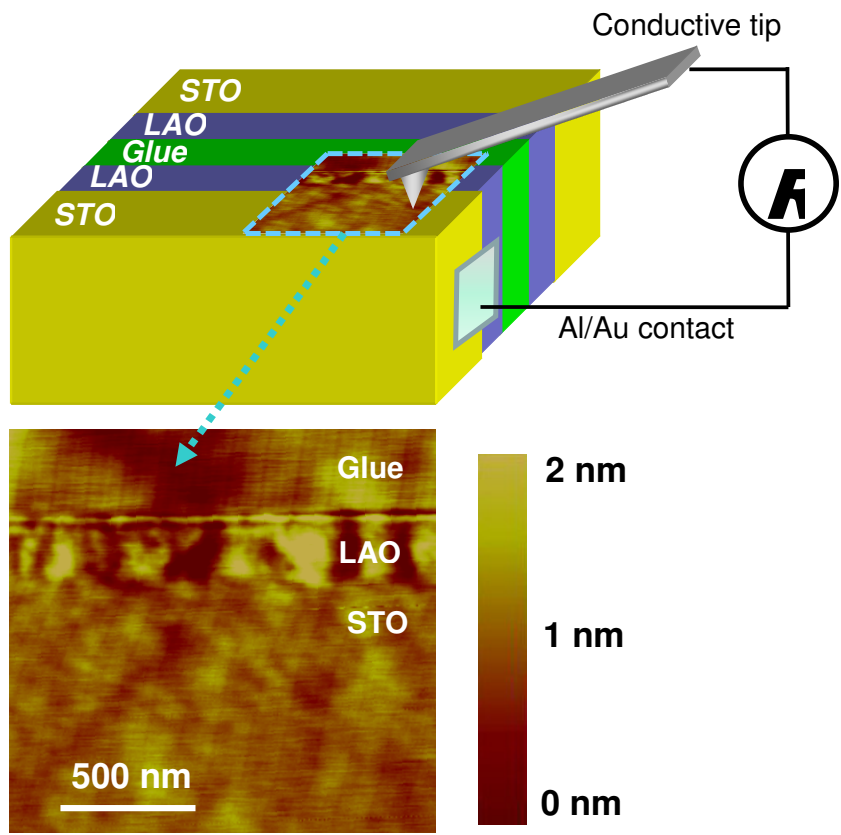

Fig. 3

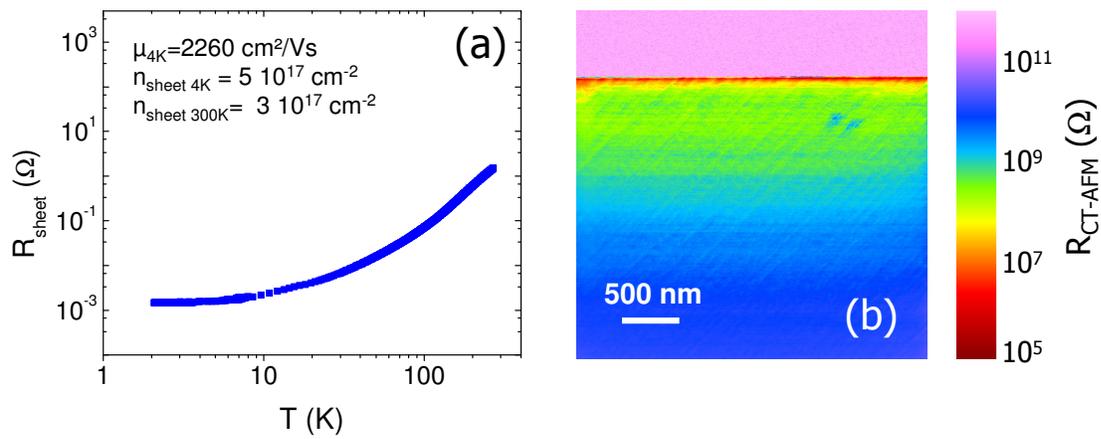

Fig. 4

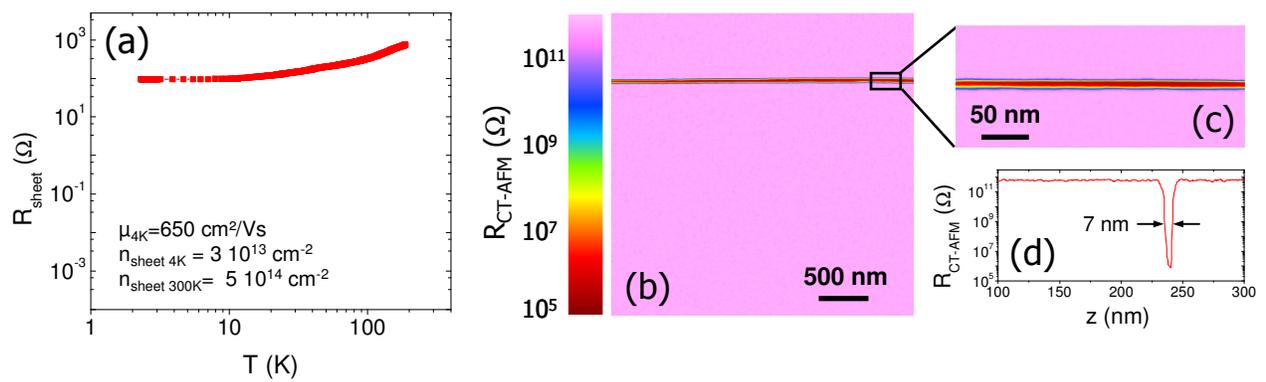

Fig. 5

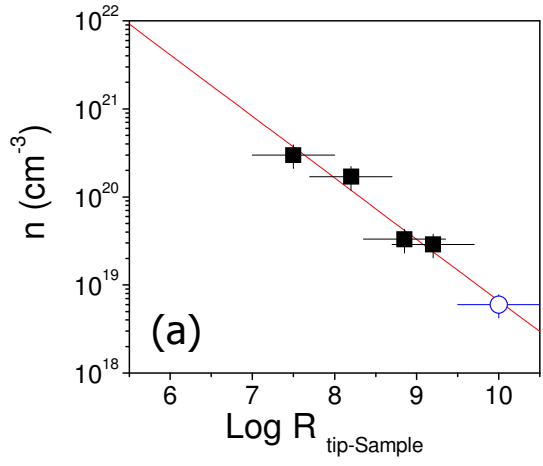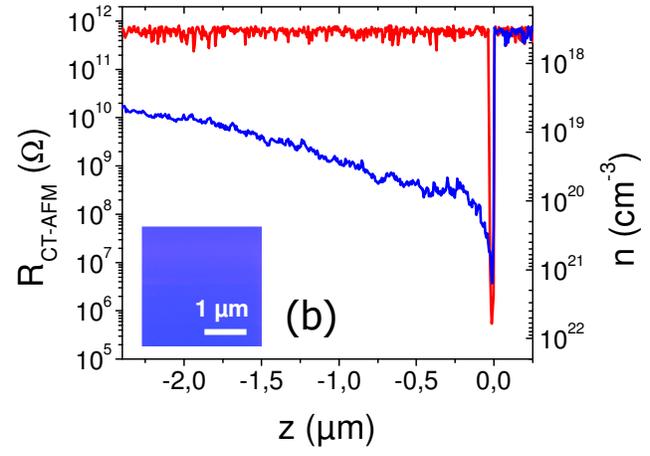

REFERENCES

1. Ohtomo, A. & Hwang, H.Y. A high-mobility electron gas at the $LaAlO_3/SrTiO_3$ heterointerface. *Nature* **427**, 423-426 (2004).

2. Pentcheva, R. & Pickett, W. E. Charge localization or itineracy at $LaAlO_3/STiO_3$ interfaces: Hole polarons, oxygen vacancies, and mobile electrons. *Phys. Rev. B* **74**, 035112 (2006)

3. Huijben, M. *et al*. Electronically coupled complementary interfaces between perovskite band insulators. *Nature Mater.* **5**, 556-560 (2006).

4. Thiel, S., Hammerl, G., Schmehl, A., Schneider, C.W. & Mannhart, J. Tunable Quasi–Two-Dimensional Electron Gases in Oxide Heterostructures. *Science* **313**, 1942-1946 (2006).

5. Herranz, G. *et al.* High Mobility in $LaAlO_3/SrTiO_3$ Heterostructures: Origin, Dimensionality, and Perspectives. *Phys. Rev. Lett.* **98**, 216803 (2007).

6. Frederikse, H. P. R. & Hosler, W. R. Hall mobility in $SrTiO_3$. *Phys. Rev.* **161**, 822-827 (1967).

7. Koonce, C. S., Cohen, M. L., Schooley, J. F., Hosler, W. R. & Pfeiffer, E. R. Superconducting transition temperatures of semiconducting $SrTiO_3$. *Phys. Rev.* **163**, 380-390 (1967).

8. Bouzehouane, K. *et al*. Enhanced dielectric properties of $SrTiO_3$ epitaxial thin films for tunable microwave devices. *Appl. Phys. Lett.* **80**, 109-111 (2002).

9. Ueda, K., Tabata, H. & Kawai, T. Ferromagnetism in $LaFeO_3/LaCrO_3$ superlattices. *Science* **280**, 1064-1066 (1998).

10. Tsukazaki, A. *et al*. Quantum Hall effect in polar oxide heterostructures. *Science* **315**, 1388-1391 (2007).

11. Nakagawa, N., Hwang, H.Y. & Muller, D. A. Why Some Interfaces Cannot be Sharp. *Nature Mater.* **5**, 204-209 (2006).

12. Siemons, W.*et al*., Origin of Charge Density at $LaAlO_3$ on $SrTiO_3$ Heterointerfaces: Possibility of Intrinsic Doping. *Phys. Rev. Lett.* **98**, 196802 (2007).

13. Park, M. S., Rhim, S. H. & Freeman, A. J. Charge compensation and mixed valency in $LaAlO_3/SrTiO_3$ heterointerfaces studied by the FLAPW method. *Phys. Rev. B* **74**, 205416 (2006).

14. Maurice, J.-L., Carrétéro, C., Casanove, M.-J., Bouzehouane, K., Guyard, S., Larquet, É. & Contour, J.-P. Electronic conductivity and structural distortion at the interface between insulators $SrTiO_3$ and $LaAlO_3$. *Phys. Stat. Sol. (a)* **203**, 2209-2214 (2006).

15. El Kazzi, M. *et al.*, Photoemission (XPS and XPD) study of epitaxial LaAlO$_3$ film grown on SrTiO$_3$(001). *Mater. Sci. Semicond. Process.* **9**, 954-958 (2006).

16. Vonk, V. *et al*. Interface structure of SrTiO$_3$/LaAlO$_3$ at elevated temperatures studied in situ by synchrotron x rays. *Phys. Rev. B* **75**, 235417 (2007).

17. Kalabukhov, A. S. *et al.* Effect of oxygen vacancies in the SrTiO$_3$ substrate on the electrical properties of the LaAlO$_3$/SrTiO$_3$ interface. *Phys. Rev. B* **75**, 121404(R) (2007).

18. Cen. C *et al.* Nanoscale control of an interfacial metal-insulator transition at room temperature. *Nature Mater.* doi:10.1038/nmat2136

19. Reyren, N. *et al.* Superconducting Interfaces Between Insulating Oxides. *Science* **317**, 1196-1199 (2007).

20. Brinkman, A. *et al*. Magnetic effects at the interface between non-magnetic oxides. *Nature Mater.* **6**, 493-496 (2007).

21. Ishigaki, T., Yamauchi, S., Kishio, K., Mizusaki, J. & Fueki, K. Diffusion of oxide ion vacancies in perovskite-type oxides. *J. Solid State Chem.* **73**, 179-187 (1988).

22. Maurice, J.-L. *et al*. Charge imbalance at oxide interfaces: How nature deals with it, *Mater. Sci. Eng. B* **144**, 1-6 (2007)

23. Maurice J.-L. *et al.* Electron energy loss spectroscopy determination of Ti oxidation state at the (001) LaAlO3/SrTiO3 interface as a function of LaAlO3 growth conditions. *Europhys. Lett.* **82**, 17003 (2008)

24. Eckstein, J. Oxide interfaces: Watch out for the lack of oxygen. *Nature Mater.* **6**, 473-474 (2007).

25 Willmott, P. R. *et al.*, Structural basis for the conducting interface between LaAlO$_3$ and SrTiO$_3$, *Phys. Rev. Lett.,* **99** 155502 (2007).

26. Houzé, F., Meyer, R., Schneegans, O. & Boyer, L. Imaging the local electrical properties of metal surfaces by atomic force microscopy with conducting probes. *Appl. Phys. Lett.*, **69**, 1975-1977 (1996).